\documentclass[12pt]{article}
\usepackage{amsmath}
\usepackage{latexsym}
\usepackage{amssymb}
\newtheorem{lem}{Lemma}
\newtheorem{thm}{Theorem}

\makeatletter
\newenvironment{pf}{\noindent{\bf Proof}\quad}
{\leavevmode\hfill$\Box$\par\@endpetrue}
\makeatother
\def\Label#1{\label{#1}\ [\ #1\ ]\ }

\def\Tr{\mathop{\rm Tr}\nolimits}
\def\SU{\mathop{\rm SU}\nolimits}
\def\id{\mathop{\rm Id}\nolimits}

\def\GL{\mathop{\rm GL}\nolimits}
\def\SL{\mathop{\rm SL}\nolimits}

\def\complex{\mathbb{C}}

\def\aku{\vspace{2ex}}

\def\Label#1{\label{#1}\ [\ #1\ ]\ }
\def\Label{\label}
\begin{document}
\vskip 2em \begin{center}
 {\LARGE\bf 
Asymptotics of Quantum Relative Entropy \par
From Representation Theoretical Viewpoint
\par} \vskip 1.5em 
\large \lineskip .5em
Masahito Hayashi \par
Department of Mathematics, Kyoto University, Kyoto 606-8502, Japan \par 
e-mail address: masahito@kusm.kyoto-u.ac.jp \par
\end{center}
\begin{abstract}
In this paper it was proved that the quantum relative entropy $D( \sigma \| \rho )$ can be asymptotically attained by Kullback Leibler divergences of probabilities given by 
a certain sequence of POVMs.
The sequence of POVMs depends on $\rho$, but is independent of the choice of $\sigma$. 
\end{abstract}
\section{Introduction}
In classical statistical theory 
the relative entropy $ D( p \| q)$ is an information quantity which means the statistical efficiency in order to distinguish a probability measure $p$ of a 
measurable space from another probability measure $q$ of the same measurable space. 
The states correspond to measures on measurable space.
When $p,q$ are discrete probabilities, the relative entropy (called also information divergence) introduced by Kullback and Leibler is defined by \cite{1}:
\begin{eqnarray*}
D(p \| q ) := \sum_i p_i \log \frac{p_i}{q_i} .
\end{eqnarray*}
In this paper, we consider the quantum mechanical case.
Let ${\cal H}$ be a separable Hilbert space which corresponds to the 
physical system of interest.
In quantum theory 
the states of a system correspond to positive 
operators of trace one on ${\cal H}$.
(These operators are called densities.)
The quantum relative entropy of a state $\rho$ with respect to another state $\sigma$ is defined by \cite{2}:
\begin{eqnarray*}
D( \sigma \| \rho ) :=
\Tr [\sigma ( \log \sigma - \log \rho )].
\end{eqnarray*}
States are distinguished through the result of a quantum measurement 
on the system.
The most general description of a quantum measurement probability is given 
by the mathematical concept of a {\it positive operator valued measure} (POVM) 
$M = \{ M_i \}_{i=1}^{h(M)}$ \cite{4,5} which is 
a partition of the unit $\id_{\cal H}$ such that
any $M_i$ is nonnegative operator.
A POVM $M=\{ M_i \}$ on ${\cal H}$ is called {\it Projection
Valued Measure} (PVM), if any $ M_i$ is projection.
In quantum mechanics, $P^M_\rho(i)= \Tr [ M_i \rho ]$
describes the probability distribution given by a POVM $M$ 
with respect to a state $\rho$.
Then we define the quantity $D_M( \sigma \|  \rho )$ as \cite{2}:
\begin{eqnarray*}
D_M( \sigma \|  \rho ) := D( P^M_\sigma \| P^M_\rho ).
\end{eqnarray*}
Thus an information quantity we can directly access by a measurement $M$ is 
not $D( \sigma \| \rho )$ but $D( P^M_\sigma \| P^M_\rho )$.
The map $\rho \mapsto P^M_\rho$ is the dual of 
a unitpreserving completely positive map.
Therefore, we have the following by Uhlmann inequality \cite{6}: 
\begin{eqnarray}
D_M( \sigma \| \rho ) \le D( \sigma \| \rho ) \Label{uhl}.
\end{eqnarray}
The equality is attained by a certain POVM $M$ when and only when 
$\rho \sigma = \sigma \rho$. 

In this paper, we consider asymptotic attainment of 
the equality of the inequality (\ref{uhl}).
In order to answer the question we define the quantum i.i.d.-condition
which is the quantum analogue of 
the independent and identically distribution condition.
If there exist $n$ samples of the state $\rho$, the quantum state is described 
by $\rho^{\otimes n}$ defined by:
\begin{eqnarray*}
\rho^{\otimes n} :=  \underbrace{\rho \otimes \cdots \otimes \rho }_{n}
\hbox{ on } {\cal H}^{\otimes n} ,
\end{eqnarray*}
where the composite system ${\cal H}^{\otimes n}$ is defined as:
\begin{eqnarray*}
{\cal H}^{\otimes n} := \underbrace{{\cal H} \otimes \cdots \otimes {\cal H}}_{n} .
\end{eqnarray*}
In this paper, we call this condition the quantum i.i.d.-condition.
Related to the inequality (\ref{uhl}),
it is well-known that 
$D(\sigma^{\otimes n}\|\rho^{\otimes n})=nD(\sigma\|\rho)$.

Let $M_n$ be a POVM on ${\cal H}^{\otimes n}$,
then we have
\begin{align}
\frac{1}{n}
D_{M_n}( \rho^{\otimes n} \| \sigma^{\otimes n} )
\le D(\sigma\|\rho) \Label{21.3}.
\end{align}
Therefore,
we consider 
the attainment of the equality of (\ref{21.3}) in taking the limit 
of $n \to \infty$.
Hiai and Petz\cite{8} proved the following theorem
with respect to this problem.
\begin{thm}
Assume that the dimension of ${\cal H}$ is finite.
Let $\sigma_n $ be a state on ${\cal H}^{\otimes n}$.
If the sequence $\{
\frac{1}{n}D( \sigma_n \| \rho^{\otimes n})\} $ 
convergence as $n \to \infty$, then 
we have
\begin{align}
\lim_{n \to \infty}
\frac{1}{n}
D( {\cal E}_{\rho^{\otimes n }}( \sigma_n) \|\rho^{\otimes n})
=
\lim_{n \to \infty}
\frac{1}{n}
D(  \sigma_n \|\rho^{\otimes n}), \Label{21.4}
\end{align}
where ${\cal E}_{\rho^{\otimes n }}$ denotes the conditional expectation
defined in (\ref{11.4.1}) in the following section.
\end{thm}
The preceding theorem implies that
\begin{align*}
\lim_{n \to \infty}
\frac{1}{n}
D_{E({\cal E}_{\rho^{\otimes n}}( \sigma_n))
\times E(\rho^{\otimes n })}
( \sigma_n \|\rho^{\otimes n})
=
\lim_{n \to \infty}
\frac{1}{n}
D(  \sigma_n \|\rho^{\otimes n}),
\end{align*}
where the PVM $E({\cal E}_{\rho^{\otimes n}}( \sigma_n))
\times E(\rho^{\otimes n })$ is defined in the following section.
In this paper, we consider whether a sequence of PVMs satisfying (\ref{11.4})
depends on $\sigma_n$ in the case of that the
state $\sigma_n$ satisfies the quantum i.i.d.-condition i.e.
$\sigma_n = \sigma^{\otimes n}$:
\begin{eqnarray}
\frac{D_{M_n}(\sigma^{\otimes n} \| \rho^{\otimes n}) }{n} \to 
 D(\sigma \| \rho )  \hbox{ as } n \to \infty ~ \forall \sigma .\Label{11.4}
\end{eqnarray}
We will consider this problem
from a representation theoretical viewpoint.
The main theorem of this paper is the following theorem.
\begin{thm}\Label{thm01}
Let $\rho$ be a state on ${\cal H}$, then
there exists a sequence $\{ ( l_n , M_n ) \}$ of pairs of 
an integer and a measurement on $ {\cal H}^{ \otimes l_n}$
such that 
\begin{eqnarray}
\frac{D_{M_n}(\sigma^{\otimes l_n} \| \rho^{\otimes l_n}) }{l_n} \to 
 D(\sigma \| \rho )  \hbox{ as } n \to \infty ~ \forall \sigma .\Label{siki01}
\end{eqnarray}
In the finite-dimensional case, 
the convergence of (\ref{siki01}) is uniform for all $\sigma$.
\end{thm}
\section{Preliminary}
Next, we consider the relation between a PVM and a quantum relative entropy.
We put some definitions for this purpose.
A state $\rho$ is called {\it commutative} with 
a PVM $E(=\{ E_i \})$ on ${\cal H}$ 
if $\rho E_i = E_i \rho $ for any $i$.
For PVMs $E (=\{ E_i \}) ,F(=\{ F_j \})$, 
we denote $E \le  F$ if for any $i$ there exist
subsets $\{ (F/E)_i \}$ such that $E_i = \sum_{j \in (F/E)_i} F_j$.
For a state $\rho$, $E(\rho)$ denotes the spectral measure of $\rho$
which can be regarded a PVM.
The {\it conditional expectation} ${\cal E}_E$ 
with respect to a PVM $E$ is defined as:
\begin{eqnarray}
 {\cal E}_E : \rho \mapsto \sum_{i} E_i \rho E_i . \Label{11.4.1}
\end{eqnarray}
Therefore, the conditional expectation ${\cal E}_E$ is
an affine map from the set of states to themselves.
Then, the state ${\cal E}_E(\rho)$ is commutative with a PVM $E$.
For simplicity, we denote the conditional expectation
${\cal E}_{E(\rho)}$ by ${\cal E}_{\rho}$.
\begin{thm}\Label{thm1}
Let $E$ be a PVM
such that $w(E):= \sup_i \dim E_i \,< \infty$.
If states $\rho,\sigma$ are commutative with a PVM $E$ and 
a PVM $F$ satisfies that $E,E({\rho}) \le F$, then
we have 
\begin{eqnarray*}
D_{F}( \sigma \| \rho ) \le D(\sigma \| \rho ) \le
D_{F}(\sigma \| \rho ) + \log w(E).
\end{eqnarray*}
\end{thm}
Note that there exists a PVM $F$ such that $E,E(\rho) \le F$.

\begin{pf}
It is proved by Lemma \ref{lem1} and Lemma \ref{lem2}.
\end{pf}
\begin{lem}\Label{lem1}
Let $\sigma,\rho$ be states.
If a PVM $F$ satisfies that $E(\rho) \le F$, then
\begin{eqnarray}
D(\sigma \| \rho ) 
= D_{F}( \sigma \| \rho ) + 
D ( \sigma \| {\cal E}_{F}(\sigma ) ) . \Label{pita}
\end{eqnarray}
\end{lem}
\begin{pf}
Since $E(\rho) \le F$,
$F$ is commutative with $\rho$,
we have $\Tr {\cal E}_{F}(\sigma)\log \rho = \Tr \sigma \log \rho$.
Remark that $\Tr {\cal E}_{F}(\sigma)\log \sigma = \Tr \sigma \log \sigma$.
Therefore, we get the following:
\begin{eqnarray*}
D_{F}( \sigma \| \rho )- D(\sigma \| \rho ) 
&=& \Tr {\cal E}_{F}(\sigma) (\log {\cal E}_{F}(\sigma) 
- \log \rho) - \Tr \sigma (
\log \sigma - \log \rho) \\
&=& \Tr {\cal E}_{F}(\sigma) (\log {\cal E}_{F}(\sigma) -\log \sigma ).
\end{eqnarray*}
We get (\ref{pita}).
\end{pf} 
\begin{lem}\Label{lem2}
Let $E,F$ be PVMs such that $E \le F$.
If a state $\sigma$ is commutative with $E$, then we have
\begin{eqnarray}
D ( \sigma \| {\cal E}_{F}(\sigma) )
\le \log w(E) . \Label{5}
\end{eqnarray} 
\end{lem}
\begin{pf}
Let 
$a_i :=\Tr E_i \sigma E_i ,\sigma_i: = \frac{1}{a_i} E_i \sigma E_i$,
then
$\sigma = \sum_{i}a_i \sigma_i$,
${\cal E}_{F}(\sigma)=
\sum_i a_i {\cal E}_{F}(\sigma_i)$.
Therefore,
\begin{eqnarray*}
&& D ( \sigma \| {\cal E}_{F}(\sigma) )
= \sum_{i} a_i D ( \sigma_i \| {\cal E}_{F}(\sigma_i) )
\le \sup_i D ( \sigma_i \| {\cal E}_{F}(\sigma_i) ) \\
&=& \sup_i \left(
\Tr \sigma_i \log \sigma_i -
\Tr {\cal E}_{F}(\sigma_i) \log {\cal E}_{F}(\sigma_i)\right) \\
&\le& - \sup_i \Tr {\cal E}_{F}(\sigma_i) \log {\cal E}_{F}(\sigma_i)
\le  \sup_i \log \dim E_i
= \log w(E).
\end{eqnarray*}
Thus, we get (\ref{5})
\end{pf}
If a PVM $F=\{ F_j \}$ is commutative with a PVM $E=\{ E_i \}$, then
we can define the PVM $F \times E= \{ F_j E_i \}$.
Then we have $F \times E \ge E,F$.
If $E'$ is commutative with $E,F$ and $F \ge E$, then
we have $E' \times F \ge E' \times E$.
If $F \ge E$ and 
$\frac{\Tr [F_j \rho] }{\Tr [E_i \rho]}
=\frac{\Tr [F_j \sigma] }{\Tr [E_i \sigma]}$
for $j \in (F/E)_i$,
then we have $D_F( \sigma \| \rho)= D_E( \sigma \| \rho)$. 
\section{Quantum i.i.d. condition from group theoretical viewpoint}\Label{s4}
From the orthogonal direct sum decomposition 
${\cal H}= {\cal H}_1 \oplus \cdots \oplus {\cal H}_k$, 
we can naturally constitute the PVM $\{ P_{{\cal H}_i} \}$,
where $P_{{\cal H}_i}$ denotes the projection of ${\cal H}_i$.
In the following, we consider the relation between 
irreducible representations and PVMs.
\subsection{group representation and its irreducible decomposition}
Let $V$ be a finite dimensional vector space 
over the complex numbers $\complex$.
A map $\pi$ from a group $G$ to the generalized linear group
of a vector space $V$ is called a {\it representation}
if the map $\pi$ is homomorphism i.e. 
$\pi( g_1 ) \pi (g_2) = \pi ( g_1 g_2 ), ~ \forall g_1 , g_2 \in G$.
For a subspace $W$ of $V$, it is {\it invariant} with respect to 
a representation $\pi$ if 
$\pi|_W (g_1) \pi|_W (g_2)= \pi|_W(g_1 g_2) ,~ \forall g_1 , g_2 \in G$,
where $\pi|_W$ denotes the restriction of $\pi$ to $W$.
In this case, $\pi|_W$ is called a {\it subrepresentation} of $\pi$.
Let $\pi$ be a representation to $V$, then
$\pi$ is called {\it irreducible} if
there no proper nonzero invariant subspace of $V$.
Let $\pi_1$($\pi_2$) be representations of a group $G$ on 
$V_1$($V_2$) respectively.
The {\it tensored} representation $\pi_1 \otimes \pi_2$ of $G$ on $V_1 \otimes V_2$
is defined as:
$(\pi_1 \otimes \pi_2) (g) 
= \pi_1  (g) \otimes \pi_2 (g) $,
and the {\it direct sum}
 representation $\pi_1 \oplus \pi_2$ of $G$ on $V_1 \oplus
 V_2$ is also defined as:
$(\pi_1 \oplus \pi_2) (g) 
= \pi_1  (g) \oplus \pi_2 (g) $.
If there is a invertible linear map $f$ from $V_1$ to $V_2$ 
such that $ f \pi_1(g) = \pi_2(g) f$,
$\pi_1$ is {\it equivalent} with $\pi_2$.
If $W$ is an invariant subspace for a representation $\pi$ on $V$ ,
then there is a complementary invariant subspace $W'$ for 
a representation $\pi$, so that $ V = W \oplus W'$ and 
$\pi= \pi|_W \oplus \pi|_{W'}$.
Therefore, any representation is a direct sum representation
of irreducible representations.

Let $\pi_1$ ($\pi_2$) be a representation of $W_1$ ($W_2$) respectively.
$W_1 \oplus W_2$ gives an irreducible decomposition of 
the direct sum representation $\pi:=\pi_1 \oplus \pi_2$.
If $\pi_1$ is equivalent with $\pi_2$, 
there is another irreducible decomposition.
For example, there is an irreducible decomposition $ 
\{ v \oplus f(v) | v \in W_1 \} \oplus 
\{ v \oplus -f(v) | v \in W_1 \}$,
where $f$ is a map which gives the equivalence with
$\pi_1$ and $\pi_2$.
If  $\pi_1$ isn't equivalent with $\pi_2$, 
there is no irreducible decomposition
except $W_1 \oplus W_2$.
A direct sum decomposition 
$W= W_1 \oplus \cdots \oplus W_s$ is called {\it isotypic} with respect to
a representation $\pi$ 
if it satisfies the following conditions:
{\it every irreducible component of $W_i$ with respect to 
a representation $\pi|_{W_i}$ is equivalent with each other.
If $i \neq j$, then any irreducible component of $W_i$ with respect to 
a representation $\pi|_{W_i}$ isn't equivalent with 
any irreducible component of $W_j$ with respect to 
a representation $\pi|_{W_j}$.}

For a representation $\pi$ of $G$,
we can define the subrepresentation $\pi|^{G_1}$ of a subgroup $G_1$ of $G$ 
by restricting a representation $\pi$ to $G_1$.
If the subrepresentation $\pi|^{G_1}$ is irreducible, then
the representation $\pi$ is irreducible.
But, the converse isn't true.
In this paper, we call a subgroup $G_1$ of $G$ {\it unramified} 
if any subrepresentation $\pi|^G$ is irreducible when the representation 
$\pi$ of $G$ is irreducible.
\subsection{Relation between a unitary representation and a PVM}
Let ${\cal H}$ be a finite-dimensional Hilbert space.
A representation $\pi$ to a Hilbert space ${\cal H}$ is called {\it unitary}
if $\pi (g)$ is a unitary matrix for any $g \in G$.
If ${\cal H}_1$ is an invariant subspace of ${\cal H}$ 
with respect to a unitary representation $\pi$,
the orthogonal space ${\cal H}_1$ of ${\cal H}_2$ is invariant 
with respect to a unitary representation $\pi$.
Therefore, we have $\pi= \pi|_{{\cal H}_1} \oplus \pi|_{{\cal H}_2}$.
A unitary representation $\pi$ can be
described by the orthogonal direct sum representation 
of irreducible representations which are orthogonal with one another.
We can regarded the direct sum decomposition as a PVM.
Remark that without unitarity we cannot deduce the orthogonality.
If there is a pair of irreducible component whose elements
are equivalent with one another.
Therefore, a corresponding PVM is not unique.
In this paper, we denote the set of PVMs corresponding to
an orthogonal irreducible decomposition by ${\cal M}( \pi )$.

Elements of the isotypic decomposition of a unitary representation $\pi$
are orthogonal with one another.
Thus, we can define a PVM $N( \pi )$ as the isotypic decomposition.
We call a representation $\pi$ of a group $G$ {\it quasi-unitary}
if there exist an unramified subgroup $G_1$ such that 
the subrepresentation $\pi|^{G_1}$ is unitary.
For a quasi-unitary representation $\pi$,
we define $N( \pi)$ (${\cal M}( \pi )$)
by $N( \pi|^{G_1} )$ (${\cal M}( \pi|^{G_1} )$) respectively.
We can show the uniqueness of them.
For a unitary representation $\pi$ and $g \in G$,
$\pi(g)$ is commutative with a PVM $M \in {\cal M}(\pi)$
and a PVM $N(\pi)$.
Concerning a quasi-unitary representation $\pi$,
we can prove the same fact.
\subsection{Relation between the tensored representation and PVMs}
Let the dimension of the Hilbert space ${\cal H}$ is $k$.
Irreducible representations of the special linear group $\SL({\cal H})$
and the special unitary group $\SU({\cal H})$
are classified by the highest weight.
Thus, any irreducible 
representation of the special linear group $\SL({\cal H})$
is irreducible under restricting 
to the special unitary group $\SU({\cal H})$.
The special unitary group $\SU({\cal H})$ is unramified subgroup
of the special linear group $\SL({\cal H})$.
Also, the special linear group $\SL({\cal H})$ is unramified subgroup
of the general linear group $\GL({\cal H})$
since the general linear group $\GL({\cal H})$ is described 
as the direct sum group $\SL( {\cal H}) \times U(1)$.

We denote the natural representation of 
the general linear group $\GL({\cal H})$,
the special linear group $\SL({\cal H})$, 
the special unitary group $\SU({\cal H})$
 to ${\cal H}$
by $\pi_{\GL({\cal H})}$, $\pi_{\SL({\cal H})}$, $\pi_{\SU({\cal H})}$,
respectively.
We consider 
representations $\pi_{\GL({\cal H})}^{\otimes n}
:= ( ( \cdots ( \pi_{\GL({\cal H})} \otimes \pi_{\GL({\cal H})}) \cdots ) 
\otimes \pi_{\GL({\cal H})})$, 
$\pi_{\SL({\cal H})}^{\otimes n}
:= ( ( \cdots ( \pi_{\SL({\cal H})} \otimes \pi_{\SL({\cal H})}) \cdots ) 
\otimes \pi_{\SL({\cal H})})$ 
and
$\pi_{\SU({\cal H})}^{\otimes n}
:= ( ( \cdots ( \pi_{\SU({\cal H})} \otimes \pi_{\SU({\cal H})}) \cdots ) 
\otimes \pi_{\SU({\cal H})})$
to the tensored ${\cal H}^{\otimes n}$.
Remark that 
$\pi_{\GL({\cal H})}^{\otimes n}|^{\SL({\cal H})}
=\pi_{\SL({\cal H})}^{\otimes n}$,
$\pi_{\GL({\cal H})}^{\otimes n}|^{\SU({\cal H})}
=\pi_{\SU({\cal H})}^{\otimes n}$.
From the unitarity of the representation $\pi_{\SU({\cal H})}^{\otimes n}$,
representations $\pi_{\GL({\cal H})}^{\otimes n}$ and
$\pi_{\SL({\cal H})}^{\otimes n}$ are quasi-unitary.
Therefore, the set ${\cal M}(\pi_{\SU({\cal H})}^{\otimes n})$ 
(the PVM $N(\pi_{\SU({\cal H})}^{\otimes n})$) 
is consistent with the sets ${\cal M}(\pi_{\SL({\cal H})}^{\otimes n}),
{\cal M}(\pi_{\GL({\cal H})}^{\otimes n})$ 
(the PVMs 
$N(\pi_{\SL({\cal H})}^{\otimes n}),N(\pi_{\GL({\cal H})}^{\otimes n})$)
and we denote it by $Ir^{\otimes n}$ ($IR^{\otimes n}$)
respectively.
      
From the Weyl's dimension formula 
((7.1.8) or (7.1.17) in Goodman-Wallch\cite{GW}),
The $n$-th symmetric space is
the irreducible subspace in the representation 
$\pi_{\GL({\cal H})}^{\otimes n}$ whose dimension is maximum.
Its dimension is the repeated combination $~_{k}H_n$
evaluated by 
$~_kH_{n} =  {n+k-1 \choose k-1} = {n+k-1 \choose n}  
=~_{n+1}H_{k-1}\le (n+1)^{k-1} $.
For $ M \in Ir^{\otimes n}$, we have the following:
\begin{align}
w( M ) \le (n+1)^{k-1} \Label{s4.2}.
\end{align}
\begin{lem}\Label{thm2}
Let $\sigma$ be a state on ${\cal H}$.
Then a PVM $M \in Ir^{\otimes n}$ and the PVM $IR^{\otimes n}$
is commutative with tensored state $\sigma^{\otimes n}$.
\end{lem}
\begin{pf}
If $\sigma \in \GL ({\cal H})$,
then this lemma is trivial.
If $\sigma \notin \GL ({\cal H})$,
there exists a sequence $\{ \sigma_i \}_{i=1}^{\infty}$ of elements
of $\GL ({\cal H})$ such that 
$\sigma_i \to \sigma$ as $i \to \infty$.
Therefore we have 
$\sigma_i^{\otimes n} \to \sigma^{\otimes n}$ as $i \to \infty$.
Because a PVM $M$ is commutative with $\sigma_i^{\otimes n}$,
the PVM $M$ is commutative with $\sigma^{\otimes n}$.
Similarly, we can prove that the PVM $IR^{\otimes n}$
is commutative with $\sigma^{\otimes n}$.
\end{pf}
From the definition of $Ir^{\otimes n}$ and $IR^{\otimes n}$,
if $j \in ( M/ IR^{\otimes n})_i$, we have 
\begin{align*}
\# ( M/ IR^{\otimes n})_i
\Tr M_j E( \rho^{\otimes n})_k \sigma^{\otimes n}
= 
\Tr IR^{\otimes n}_i  E( \rho^{\otimes n})_k \sigma^{\otimes n}, 
\end{align*}
for states $\rho,\sigma$ and a PVM $M \in Ir^{\otimes n}$.
The number $\# ( M/ IR^{\otimes n})_i$ 
corresponds to the number of equivalent irreducible representations
in the representation $\pi_{\GL({\cal H})}^{\otimes n}$.
Therefore we obtain 
\begin{align}
D_{IR^{\otimes n}\times E( \rho^{\otimes n})} 
( \sigma^{\otimes n} \| \rho^{\otimes n})
= 
D_{M\times E( \rho^{\otimes n})} 
( \sigma^{\otimes n} \| \rho^{\otimes n}). \Label{12.1}
\end{align}
\section{Proof of Main Theorem}
First we will prove Theorem \ref{thm01} in the case of that the dimension
of ${\cal H}$ is finite number $k$.
Let $\rho$ be a state on ${\cal H}$.
From Theorem \ref{thm1}, Lemma \ref{thm2} and the preceding discussion,
we obtain the following fact.
For a PVM $E^n \in Ir^{\otimes n}$,
the PVM $M^n:= E^n \times E(\rho^{\otimes n})$ 
satisfies:
\begin{eqnarray}
\frac{D_{M_n}(\sigma^{\otimes n} \| \rho^{\otimes n}) }{n} \le D(\sigma \| \rho )
\le \frac{D_{M_n}(\sigma^{\otimes n} \| \rho^{\otimes n}) }{n} + (k-1)\frac{\log (n+1)}{n}
~ \forall \sigma . \Label{siki1}
\end{eqnarray}
Therefore we obtain
\begin{eqnarray}
\frac{D_{M_n}(\sigma^{\otimes n} \| \rho^{\otimes n}) }{n} \to 
 D(\sigma \| \rho )  \hbox{ as } n \to \infty \hbox{ uniformally for } \sigma .
\label{uni}
\end{eqnarray}
From (\ref{12.1}), the PVM $IR^{\otimes n} \times E(\rho^{\otimes n})$
satisfies (\ref{siki1}) and (\ref{uni}).
We get (\ref{siki01}) in the finite-dimensional case.
In spin 1/2 system,
the PVM $IR^{\otimes n}$ corresponds to 
the measurement of the total momentum, and
$E(\rho^{\otimes n})$ does to the one of
the momentum of the direction specified by $\rho$.
These obsarvables are commutative with one another.
Next, we consider the infinite-dimensional case.
Let ${\cal B}({\cal H})$ be the set of bounded operators on ${\cal H}$,
and 
${\cal B}({\cal H})^{\otimes n}$ be
$\underbrace{{\cal B}({\cal H}) \otimes 
\cdots \otimes {\cal B}({\cal H})}_{n}$.
According \cite{7},
from the separability of ${\cal H}$,
there exists a finite-dimensional approximation of ${\cal H}$.
i.e. a sequence 
$\{ \alpha_n : {\cal B}({\cal H}_n) \to {\cal B}({\cal H}) \}$
of unitpreserving completely positive maps
such that ${\cal H}_n$ is finite-dimensional
and 
\begin{eqnarray}
\lim_{n \to \infty} D( \alpha_n^*(\sigma) \| \alpha_n^*(\rho) )
= D( \sigma \| \rho ) \label{uni2}
\end{eqnarray}
for any states $\sigma , \rho$ on ${\cal H}$ 
such that $\mu \rho \le \sigma \le \lambda \rho$ for some positive real 
numbers $\mu, \lambda$.
From (\ref{uni}),
for any positive integer $n$ there exists a pair $( l_n , M'_n )$ of an integer and 
a PVM on $ {\cal H}_n^{\otimes l_n}$ such that
\begin{eqnarray}
 D( \alpha_n^*(\sigma) \| \alpha_n^*(\rho) )
- \frac{D_{M'_n}
\left( (\alpha_n^*(\sigma))^{\otimes l_n} \left\| 
(\alpha_n^*(\rho))^{\otimes l_n} \right. \right) }
{l_n}
\,< \frac{1}{n} \Label{uni3}.
\end{eqnarray}
The completely positive map $\alpha_n^{\otimes l}$
from ${\cal B}({\cal H}_n)^{\otimes l}$ to ${\cal B}({\cal H})^{\otimes l}$ is
defined as
$\alpha_n^{\otimes l}( A_1 \otimes A_2 \otimes \cdots \otimes A_l )
= \alpha_n ( A_1) \otimes  \alpha_n ( A_2) \otimes \cdots 
\otimes  \alpha_n ( A_l)$ for $\forall A_i \in {\cal B}({\cal H})$.
Therefore we have $\left(\alpha_n^{\otimes l}\right)^{*} (\sigma^{\otimes l})=
\alpha_n^*(\sigma)^{\otimes l}$.
Let $M_n$ be $\alpha_n^{\otimes l_n}(M_n')$, 
then from (\ref{uni2})(\ref{uni3}) we get
\begin{eqnarray*}
\frac{D_{M_n}(\sigma^{\otimes l_n} \| \rho^{\otimes l_n}) }{l_n} 
&=& \frac{D_{M_n'} \left( \left. \left(\alpha_n^{\otimes l_n}\right)^{*}
\left(\sigma^{\otimes l_n} \right) \right\| 
\left(\alpha_n^{\otimes l_n}\right)^{*}
\left(\rho^{\otimes l_n}\right) \right) }{l_n}  \\
&=& \frac{D_{M'_n}\left( \left. \alpha_n^*
(\sigma)^{\otimes l_n} \right\| \alpha_n^*(\rho)^{\otimes l_n} \right)}
{l_n} \\
&\,>& D \left( \alpha_n^*(\sigma) \| \alpha_n^*(\rho) \right) + \frac{1}{n} \\
&\to&  D (  \sigma \| \rho ) \hbox{ as } n \to \infty.
\end{eqnarray*}
Therefore, we obtain Theorem \ref{thm01}.
Note that such a POVM $M_n$ is independent of $\sigma$.

\section*{Conclusions}
It was proved that the quantum relative entropy $D(\sigma \| \rho )$ is attained by the sequence of Kullback-Leibler divergences given by 
 a certain sequence of POVMs which is independent of $\sigma$.
This formula is thought to be important for the quantum asymptotic detection and 
the quantum asymptotic estimation.
About the quantum asymptotic estimation, see \cite{10}.
The realization of the sequence of measurements are left for future study.
In spin 1/2 system,
it is enough to simultaneously measure the 
total momentum and the momentum of the direction specified by $\rho$.

\section*{Acknowledgments}
The author wishes to thank Prof. D. Petz for essential suggestions about 
the extension in the infinite-dimensional case. 
He wishes to thank Dr. A. Fujiwara and Dr. M. Ishii for several discussions on this topic.

\end{document}